\documentclass[conference]{IEEEtran}
\IEEEoverridecommandlockouts
\usepackage{cite}
\usepackage{amsmath,amssymb,amsfonts}
\usepackage{algorithmic}
\usepackage{graphicx}
\usepackage{booktabs}
\usepackage{array}
\usepackage{tabularx}
\usepackage{float}
\usepackage{textcomp}
\usepackage{xcolor}
\usepackage{makecell}
\usepackage{caption}
\def\BibTeX{{\rm B\kern-.05em{\sc i\kern-.025em b}\kern-.08em
    T\kern-.1667em\lower.7ex\hbox{E}\kern-.125emX}}
\begin{document}
%Using Linear Feature Decoupling method to Enhance the Fitting Accuracy of Deep learning networks for NLSE 
\title{Improve the Fitting Accuracy of Deep Learning for the Nonlinear Schrödinger Equation Using Linear Feature Decoupling Method  \\
% \title{Enhanced NLSE fitting method based feature decoupling method\\
}
%我们利用分布式特征解耦（FDD）方法来增强深度学习方法对非线性薛定谔方程（NLSE）的拟合程度，与非解耦模型相比，FDD模型的NLSE损失显著降低。
\author{
    \IEEEauthorblockN{Yunfan Zhang, Zekun Niu, Minghui Shi, Weisheng Hu, Lilin Yi*}
    \IEEEauthorblockA{
        \textit{State Key Lab of Advanced Communication Systems and Networks,}\\
        \textit{School of Electronic Information and Electrical Engineering}\\
        \textit{Shanghai Jiao Tong University}\\
        Shanghai, China\\
        e-mail: lilinyi@sjtu.edu.cn
    }
}

% \and
% \and
% \and
% \and
% \and

\maketitle

\begin{abstract}
We utilize the Feature Decoupling Distributed (FDD) method to enhance the capability of deep learning to fit the Nonlinear Schrödinger Equation (NLSE), significantly reducing the NLSE loss compared to non decoupling model.
% We analyze and utilize the Feature Decoupling Distributed Modeling (FDD) method to simplify and improve the fitting degree of neural network system to the nonlinear Schrödinger equation. Simulation experiments have verified that FDD can significantly improve the fitting degree of the model to the NLSE.

\end{abstract}
\vspace{+1em} % 根据需要调整这个值
\vspace{-1em} 
\section{Introduction}
The Nonlinear Schrödinger Equation (NLSE) describes propagation of optical pulses through optical fiber channel, which is the cornerstone for studying nonlinear optics and optical fiber communication[1][2]. The NLSE is non-analytical when both linearity and nonlinearity are considered. Traditional method to solve NLSE is based on Split-Step Fourier method (SSFM)[3], which is a numerical method and needs large iteration, leading to high computational complexity and making it challenging to meet practical engineering application[9]. 

Recently, deep learning (DL) has emerged as a accurate and efficient approach for solving NLSE and modeling optical fiber channel, owing to its their strong nonlinear fitting capabilities and efficiency in parallel computation. Neural networks, such as BiLSTM, CGAN and Transformer have succeed in modeling optical fiber channel in signal- and multi-channel WDM systems[6][7][10]. Physics informed neural networks has achieved modeling of fiber channels or optical pulses under specific boundary conditions[4][5]. 

However, most existing DL-based schemes utilize pure-data driven or incorporate physical prior knowledge in the loss function. NNs need to learn all linear and nonlinear coupled characteristic from whole NLSE, which face accuracy degrade in more complex scenarios. Meanwhile, NNs may have learned only the features of the training data, not the mappings that satisfy the NLSE equation.

In this paper, we utilize a feature decoupling distributed (FDD) method to fit NLSE in optical fiber communication.This method decouple the linear features by combining prior physical models. By embedding parameters and differentiating signal waves, the NLSE loss is assessed during training, allowing for further performance analysis. The FDD method demonstrates a threefold reduction in the predicted waveform Normalized Mean Square Error (NMSE) during training scenarios compared to the non-decoupled method. Furthermore, in non-training scenarios, the waveform NMSE is reduced by more than tenfold. For transmission distances ranging from 10-100 km, the NMSE accuracy of the FDD model remains below 
 , with a two-order magnitude reduction in NLSE loss. These results demonstrates that the FDD model achieves superior accuracy over the non-decoupling model and adapts effectively across a broader range of boundary conditions.

\begin{figure*}[t] % [t] 表示将图片放置在页面顶部
    \centering
    \includegraphics[width=\textwidth]{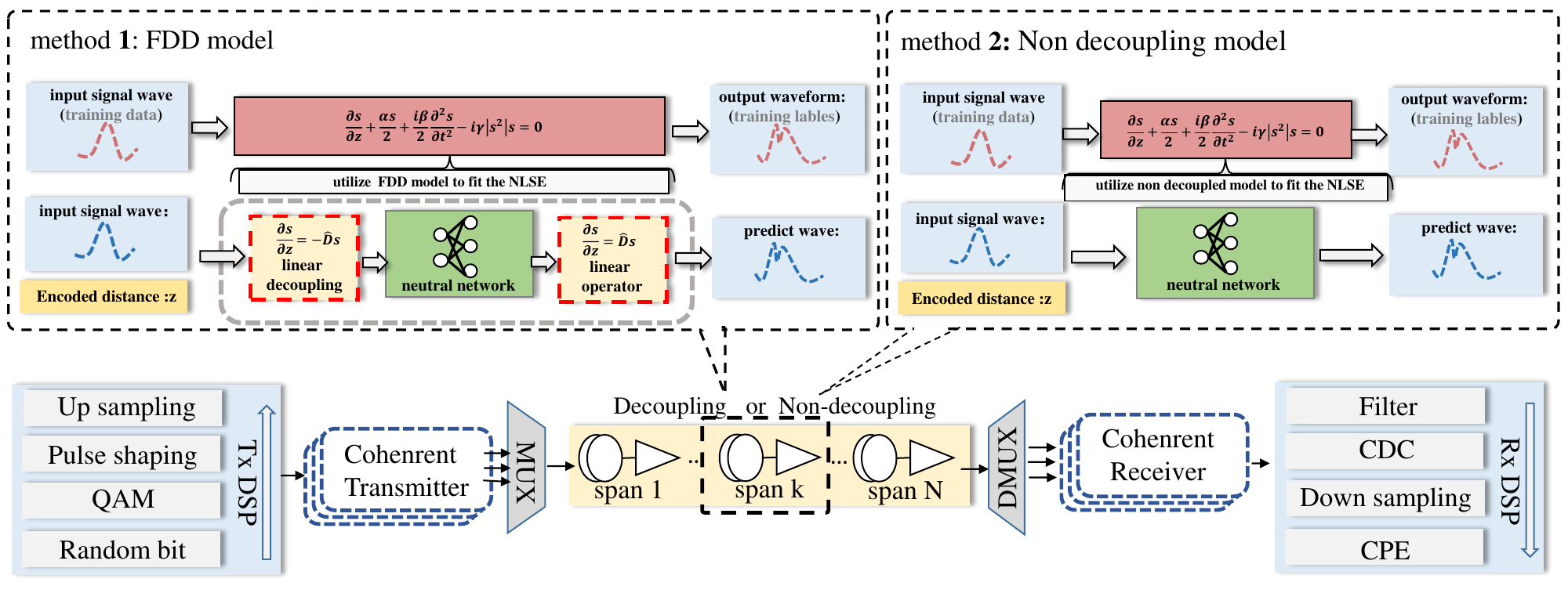} % 替换 fig1.pdf 为你的图片文件名
    \caption{Two methods for fitting the NLSE.}
    \label{fig:fig1}
\end{figure*}
\section{Principle}

\subsection{The Nonlinear Schrödinger Equation (NLSE)}
 The Nonlinear Schrödinger Equation (NLSE) is a type of Partial Differential Equations (PDEs), it describes the dispersion, nonlinear effects, and gain-loss experienced by the optical pulse as it travels through the fiber. Considering the Polarization Mode Dispersion (PMD) effect, its general form is:
\begin{equation}
\begin{split}
\frac{\partial s}{\partial z} + \frac{\alpha s}{2} + \frac{i \beta}{2} \frac{\partial^2 s}{\partial t^2} - \frac{8}{9}i \gamma |s|^2 s = 0,
\end{split}
\end{equation}
%\vspace{+0.5em} % 根据需要调整这个值
where \( s \)\( = \)  \( s(z,t) \) represents the optical field envelope, 
\( z \) is the propagation distance, 
 \( t \) is time, 
\( \beta_2 \)
  is the dispersion coefficient, 
 \( \alpha \)  is the attenuation coefficient, and 
\( \gamma \) is the nonlinear coefficient. Traditional methods for solving the NLSE use numerical simulations. The most commonly used method for solving equation (1) is the Split-Step Fourier Method (SSFM) [1]. NLSE can be written in the form of linear and nonlinear operators for signal operations, as shown in equatuin(2),
\begin{equation}
\begin{split}
\frac{\partial S(z,t)}{\partial z} = \left( \hat{D} + \hat{N}[S(z,t)] \right) S(z,t),
\end{split}
\end{equation}
where \( \hat{D} \) is the linear component, which denotes the effects
of attenuation and CD, and \( \hat{N} \) is the nonlinear component, \( S \) denotes the optical field complex envelope,
\( z \) represents the distance, and \( t \) is the time, so the SSFM treats the fiber channel propagation system as an alternating iteration of nonlinear and linear steps.

\subsection{Principle of utilizing FDD model to fit NLSE}
In the method of utilizing purely data-driven neural networks to simulate the NLSE, the neural network directly learns both the linear and nonlinear aspects of the entire NLSE system, as shown in Figure 1. From (2), it can be seen that the NLSE equation includes both linear and nonlinear operators. Therefore, to mitigate the linear effects after the entire transmission and introduce physical formula modeling, we adopt Feature Decoupling Distribution (FDD) modeling in the neural network part. This approach reduces the difficulty of fitting the neural network, however, the neural network itself no longer conforms to the NLSE. To allow the neural network to fully fit the NLSE, a linear system modeled using physical formulas is cascaded after the neural network. The linear  syetem corresponds to the inverse process of dispersion compensation and distance attenuation compensation before the neural network, which can be regarded as a linear operator, as shown in Equation (3),
\begin{equation}
\frac{\partial s}{\partial z} + \frac{\alpha s}{2} + \frac{i\beta}{2} \frac{\partial^2 s}{\partial t^2} = 0,
\end{equation}
where \( s \)\( = \)  \( s(z,t) \) represents the optical field envelope, 
\( z \) is the propagation distance, 
 \( t \) is time, 
\( \beta_2 \)
  is the dispersion coefficient, 
 \( \alpha \)  is the attenuation coefficient.The frequency domain analytical solution is given in Equation (4),
\begin{equation}
S(z + L_{\text{span}}, \omega) = S(z, \omega) \exp \left( - \frac{i \beta_2}{2} \omega^2 L_{\text{span}} \right),
\end{equation}
where 
 \( S(z,w) \)  is the Fourier transform of \( s(z,t) \) , and \(L_{\text{span}} \)
is the transmission distance for one span.The equivalent system formed by cascading the neural network and the linear system can fit the complete NLSE system.To derive the output waveform with respect to the distance parameter \( z \) and enhance the model's generalization over distances, we have implemented an encoded input method for \( z \) at the input end[8]. The chosen neural network is Bidirectional Long Short-Term Memory (BiLSTM), known for its capability to capture and learn the dependencies in the input signals[7].

%\subsection{Calculating the NLSE loss for non decoupling model and FDD model}

To obtain NLSE loss for non decoupling and FDD systems, we need to calculate the derivatives of the output signal with respect to the distance 
\( z \) and time \( t \) for the neural network. We introduce a parameter 
\( z \) that controls the distance encoded by the neural network at the input end[8]. We utilize the feature of neural networks that can back propagate and take derivatives, we obtain the derivative of the output predicted waveform with respect to \( z \). Due to the fact that our neural network is used for time-domain waveform simulation modeling without inputting the time parameter t at the input, the derivative of time \( t \) is obtained by directly taking the second-order derivative of the output waveform in time-domain. The output time-domain waveform is transformed to the frequency domain using the Fourier transform, and multiplied by 
\( iw \) squared in the frequency domain to obtain the second-order derivative of the system with respect to time, as shown in Equation (5). 
\vspace{-0em} % 根据需要调整这个值
\begin{equation}
\mathcal{F} \left\{ \frac{\partial^2 s}{\partial t^2} \right\} = (i \omega)^2 \mathcal{F} \{ s(z,t) \},
\end{equation}
where \(\mathcal{F} \) represents Fourier transform and \( i \) represents imaginary unit.
Since the simulation system we are using is a dual polarization system, we need to consider the Polarization Mode Dispersion (PMD) effect, by deriving the NLSE, we can obtain the fiber channel formula that takes into account the PMD effect, namely the Manakov PMD equation, 
%therefore, we need to use NLSE that takes into account PMD effects to calculate NLSE loss.
we substitute the time and distance derivatives of the signal and the obtained signal into equation (6) which shown as:
\begin{equation}
\ NLSE loss= \frac{\partial s}{\partial z} + \frac{\alpha s}{2} + \frac{i \beta}{2} \frac{\partial^2 s}{\partial t^2} - \frac{8}{9}i \gamma |s|^2 s,
\end{equation}
to obtain the NLSE loss of the system,
where \( s \) represents a signal containing two orthogonal polarization modes.

%\vspace{-0em} % 根据需要调整这个值
%\vspace{+0.5em} % 根据需要调整这个值

\section{Simulation and results discussion}
%\subsection{System Setup}
\subsection{Simulation and training setup}
\vspace{-0em} % 根据需要调整这个值
To collect training data and analyze the performance of channel modeling, we simulated a dual-polarization coherent WDM optical transmission system, as shown in Figure 1. The symbol rate is 30 Gbaud per channel, and the channel spacing is 50 GHz. The modulation format is dual-polarization 16 QAM, with 5 channels, and each channel has a transmit power of 5 dBm. The transmitter uses a root-raised cosine (RRC) filter with a roll-off factor of 0.1 for pulse shaping. To simulate the time-domain waveform, the channel employs 4x oversampling and uses the SSFM algorithm for simulation, as shown in Figure 1 .To enable the neural network to learn the characteristics of signal waveforms that change with different times and distances, optical signal data was collected at transmission distances of 10 km, 20 km, 30 km, 50 km, and 70 km. After passing through the backward dispersion system, the complex signals of the two polarization channels are arranged into a one-dimensional vector. 
%Although the backward dispersion system has eliminated most of the inter-symbol dispersion effects, there is still crosstalk. Therefore, multiple time series symbols need to be input each time to predict the output waveform. 
The transmission distance parameter \( z \)  is also encoded by a linear layer and input into a neural network.

\vspace{-1.5em} % 根据需要调整这个值

\begin{figure}[H] %表示将图片放置在此处
    \centering
    \includegraphics[width=\linewidth]{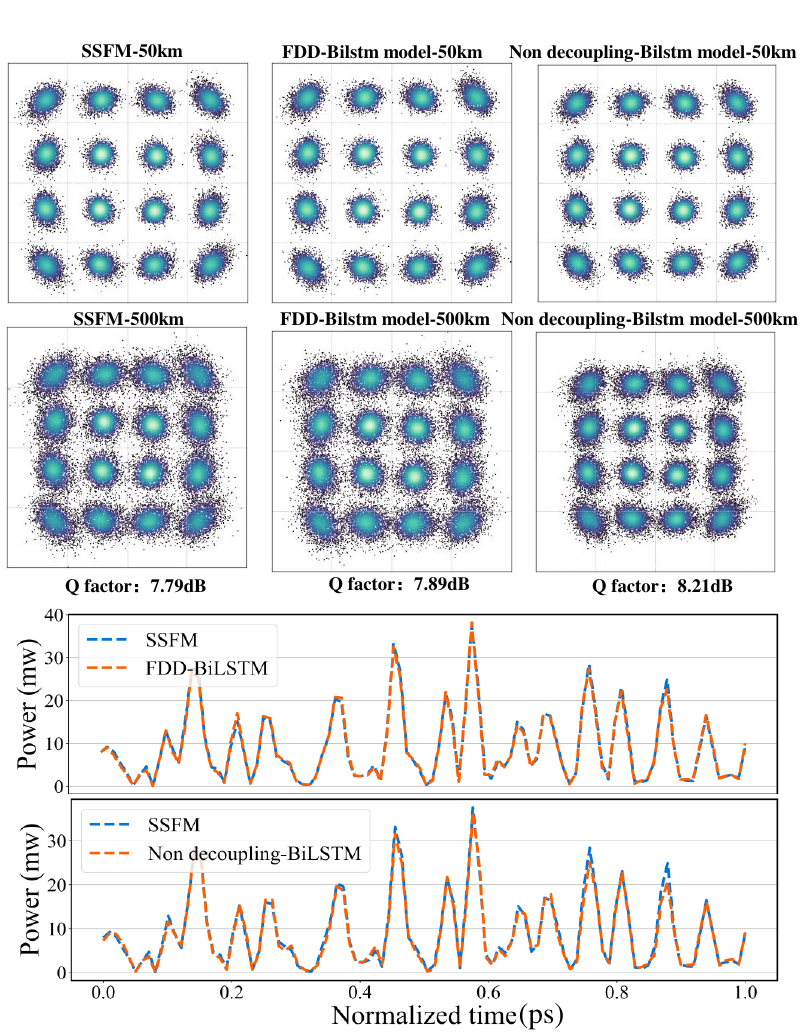} % 替换 image.png 为你的图片文件名
    \caption{The model accuracy of FDD model and non decoupling model.}
    \label{fig:single-column-image}
\end{figure}
\vspace{-0em} % 根据需要调整这个值
%\section{Results and Discussion}
\vspace{-0.5em} % 根据需要调整这个值

\subsection{Accuracy of non decoupling model and FDD model}
We compared the waveforms, constellations, Q-factors, and NMSE accuracy of the FDD model and the non decoupling model at different transmission distances to verify that the FDD model has higher waveform prediction accuracy. After DSP processing of the output waveform, as shown in Figure 2, the data points in the FDD model constellations have a rotation degree that is basically consistent with SSFM. Clearly, compared to the non decoupling model, the constellations of the FDD model exhibits consistent rotation with the constellations obtained from numerical methods, indicating that linear decoupling helps the neural network better learn the nonlinear parts of the NLSE equation. Additionally, the time-domain waveform diagram also shows that the waveforms closely match the theoretical values, which demonstrates that FDD model has higher waveform prediction accuracy.
\vspace{-1em} % 根据需要调整这个值
\begin{figure}[H]% [H] 表示将图片放置在此处
    \centering
    \includegraphics[width=\linewidth]{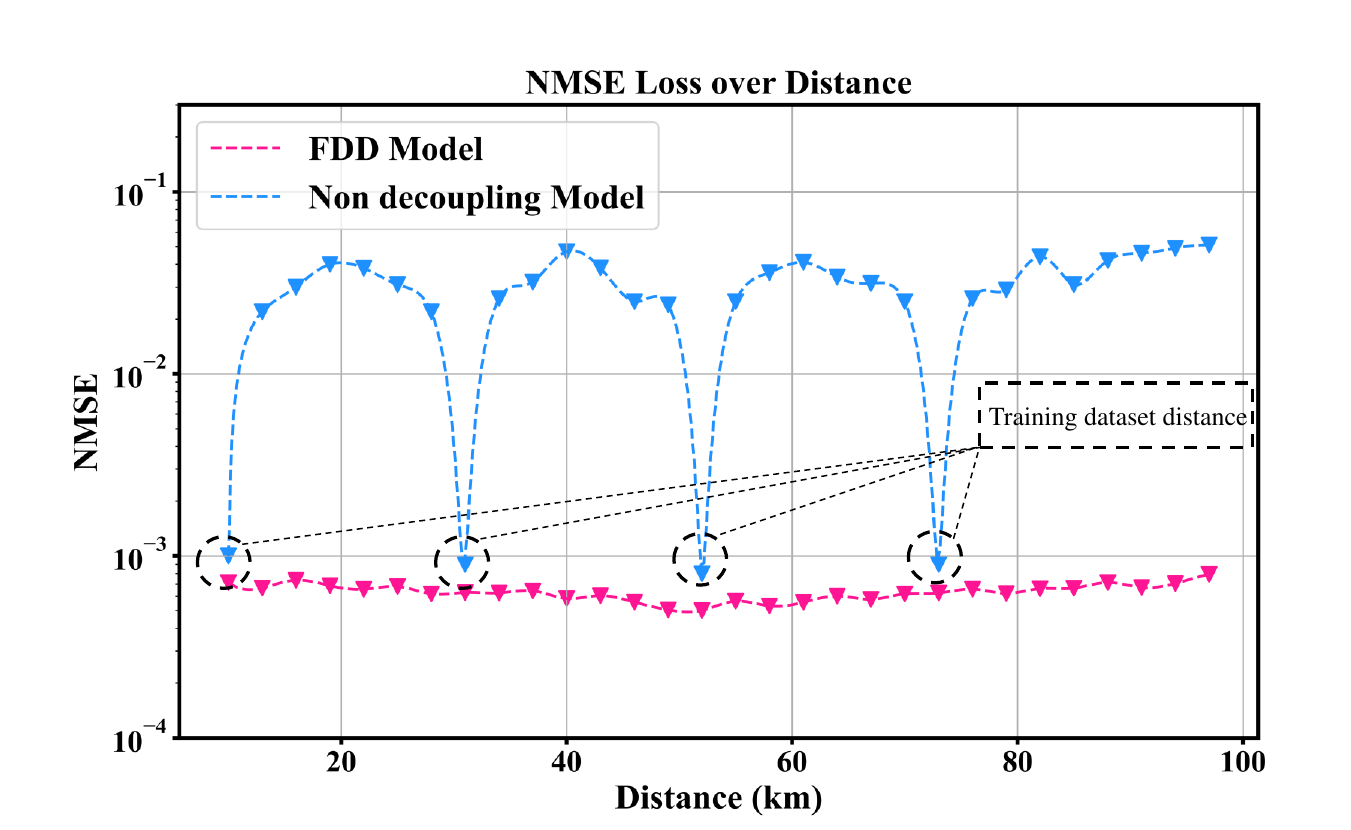} % 替换 image.png 为你的图片文件名
    \caption{NMSE of non decoupling systems and FDD model.}
    \label{fig:single-column-image}
\end{figure}
\vspace{-1em} % 根据需要调整这个值

\begin{figure}[H] % [H] 表示将图片放置在此处
    \centering
    \includegraphics[width=\linewidth]{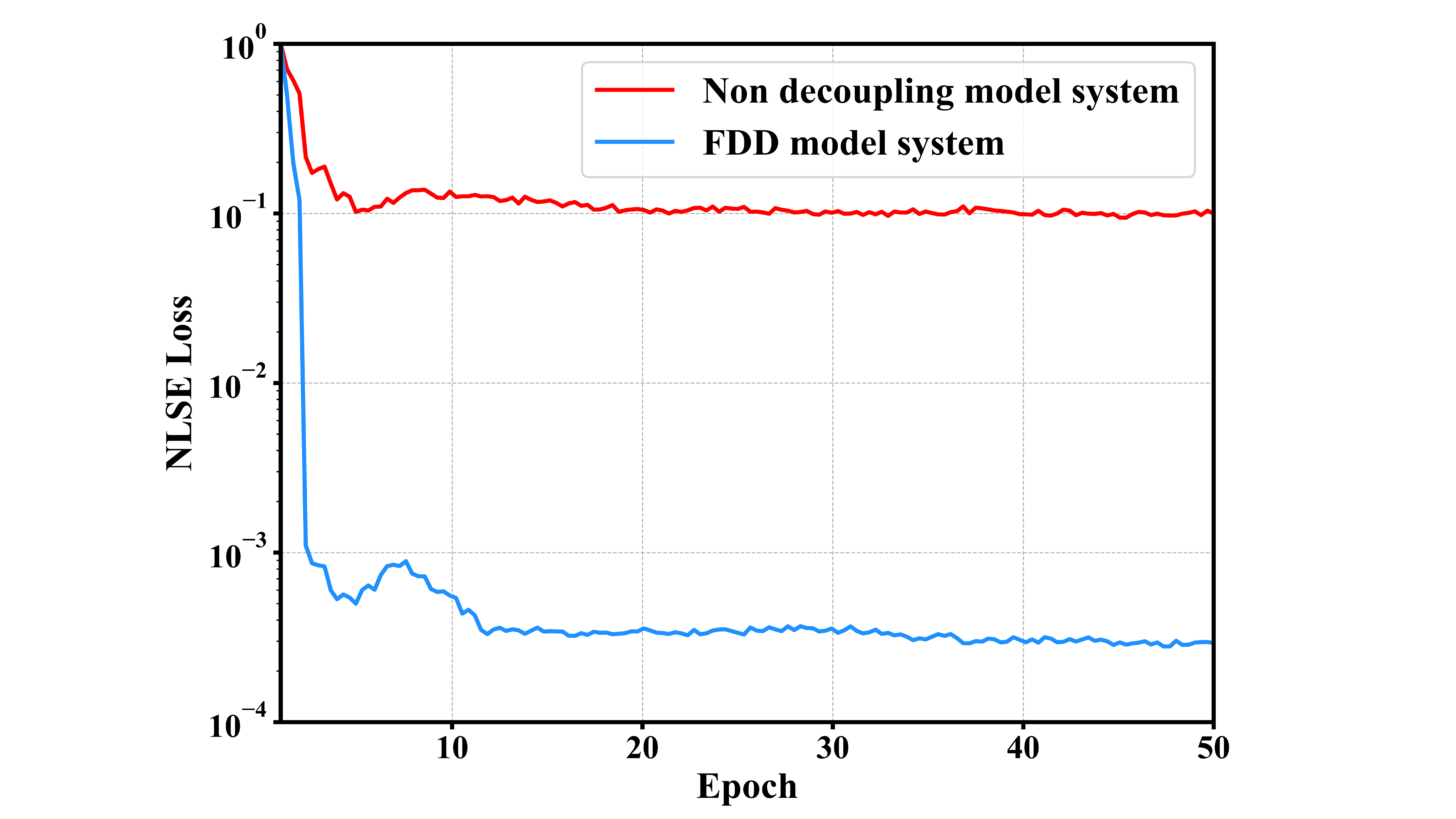} % 替换 image.png 为你的图片文件名
    \caption{NLSE loss training curve of non decoupling model and FDD model.}
    \label{fig:single-column-image}
\end{figure}
\vspace{-0.5em}
To verify that the FDD model has better generalization to distance and obtains a mapping that adapts in a wider range of boundary conditions compared to the non decoupling model, we plotted the NMSE accuracy comparison of the predicted output waveforms of the two models with different distance parameters as input. It can be seen that the waveform error accuracy of the FDD model is 1-2 orders of magnitude lower than that of the non decoupling model. 
From the figure, it can be seen that although both models adopted the training model with distance parameter \( z \) as input, the non decoupled model only has high accuracy in the distance of the training dataset, at untrained distance points, the accuracy of waveform NMSE still significantly decreases, indicating that it only learned some local optima. 
%This suggests that the non decoupling neural network model did not truly learn the NLSE function that varies with parameters 
%\( z \) and \( t \).
In contrast, the FDD model maintained high prediction accuracy for the optical signal waveform transmission over distances of 10-100 km, this indicates that the FDD model has higher accuracy in distance generalization and can obtain a mapping that adapts in a wider range of boundary conditions.

\subsection{NLSE loss of FDD model system and non decoupling model system}

To further demonstrate that the FDD model fits the NLSE equation better, we also presented the changes in NLSE loss during the training process for both the non decoupling model and the FDD model. The NLSE loss is calculated by equation (6) and the average NLSE loss of 1000 signals is taken. The NLSE loss of the non decoupling model is around the order of \(10^{-2}\), showing some decline only at the beginning of the training. This indicates that the non decoupling model clearly failing to learn the equation varying with parameters 
\( z \) and 
\( t \), but merely fitting to specific data features.
In contrast, as the number of epochs increases, the overall NLSE loss continues to gradually decrease to a lower level, with the signal amplitude around the order of and the average single-symbol NLSE loss around the order of 
\(10^{-4}\), as shown in Figure 4. In FDD model, the linear effects of the FDD model system are modeled by equations which incorporates prior physical knowledge of the linear part into the system, thus, at the beginning of the training, the linear part of the NLSE equation is already correct, allowing the neural network to focus on optimizing the nonlinear part. The lower NLSE loss explains why the FDD model has stronger distance generalization capabilities and can obtain a mapping that adapts in a wider range of boundary conditions.

\section{Conclusion}

To simplify the difficulty of fitting the NLSE equation and improve the accuracy of neural networks in fitting the NLSE equation, we applied the Feature Decoupling Distribution (FDD) method in neural networks for NLSE modeling, we analyzed and calculated the NLSE loss of both the FDD model and the non decoupling model and compared their distance generalization accuracy. Since the FDD model incorporates prior physical knowledge of the linear part of the NLSE into the system, the FDD model can better fit and approximate the original NLSE. We found that the FDD model's accuracy and distance generalization are higher than that of the non decoupling model. Additionally, the NLSE loss of the FDD model is significantly lower than that of the non decoupling system which indicates that FDD model can obtain a mapping that adapts in a wider range of boundary conditions. We believe that this method of utilizing linear feature decoupling can also be applied to neural networks that fit physical processes controlled by other Partial Differential Equations(PDEs), enabling these networks to better fit the PDEs.

% \begin{figure}[htbp]
% \centerline{\includegraphics{fig1.png}}
% \caption{Example of a figure caption.}
% \label{fig}
% \end{figure}

% Figure Labels: Use 8 point Times New Roman for Figure labels. Use words 
% rather than symbols or abbreviations when writing Figure axis labels to 
% avoid confusing the reader. As an example, write the quantity 
% ``Magnetization'', or ``Magnetization, M'', not just ``M''. If including 
% units in the label, present them within parentheses. Do not label axes only 
% with units. In the example, write ``Magnetization (A/m)'' or ``Magnetization 
% \{A[m(1)]\}'', not just ``A/m''. Do not label axes with a ratio of 
% quantities and units. For example, write ``Temperature (K)'', not 
% ``Temperature/K''.

\section*{Acknowledgment}
The authors acknowledge the funding provided by the National Key R\&D Program of China (2023YFB2905400), National Natural Science Foundation of China (62025503), and Shanghai Jiao Tong University 2030 Initiative.

% \section*{References}

% Please number citations consecutively within brackets \cite{b1}. The 
% sentence punctuation follows the bracket \cite{b2}. Refer simply to the reference 
% number, as in \cite{b3}---do not use ``Ref. \cite{b3}'' or ``reference \cite{b3}'' except at 
% the beginning of a sentence: ``Reference \cite{b3} was the first $\ldots$''

% Number footnotes separately in superscripts. Place the actual footnote at 
% the bottom of the column in which it was cited. Do not put footnotes in the 
% abstract or reference list. Use letters for table footnotes.

% Unless there are six authors or more give all authors' names; do not use 
% ``et al.''. Papers that have not been published, even if they have been 
% submitted for publication, should be cited as ``unpublished'' \cite{b4}. Papers 
% that have been accepted for publication should be cited as ``in press'' \cite{b5}. 
% Capitalize only the first word in a paper title, except for proper nouns and 
% element symbols.

% For papers published in translation journals, please give the English 
% citation first, followed by the original foreign-language citation \cite{b6}.

\vspace{12pt}
% \color{red}
% IEEE conference templates contain guidance text for composing and formatting conference papers. Please ensure that all template text is removed from your conference paper prior to submission to the conference. Failure to remove the template text from your paper may result in your paper not being published.

\end{document}